\documentclass[12pt,preprint]{aastex}
\usepackage{textcomp}
\usepackage{amsmath}
\usepackage{graphicx}

\shorttitle{The vertical composition of neutrino-dominated accretion disk}
\shortauthors{Liu et al.} \slugcomment{}

\def\beq{\begin{eqnarray}}
\def\eeq{\end{eqnarray}}

\begin{document}

\title{The vertical composition of neutrino-dominated accretion disks in gamma-ray bursts}
\author{Tong Liu\altaffilmark{1}, Li Xue\altaffilmark{1}, Wei-Min Gu\altaffilmark{1,2}, and Ju-Fu Lu\altaffilmark{1}}

\altaffiltext{1}{Department of Astronomy and Institute of Theoretical Physics and Astrophysics, Xiamen University, Xiamen, Fujian 361005, China}
\altaffiltext{2}{Harvard-Smithsonian Center for Astrophysics, 60 Garden Street, Cambridge, MA 02138, USA}
\email{tongliu@xmu.edu.cn}

\begin{abstract}
We investigate the vertical structure and elements distribution of neutrino-dominated accretion flows around black holes in spherical coordinates with the reasonable nuclear statistical equilibrium. According our calculations, heavy nuclei tend to be produced in a thin region near the disk surface, whose mass fractions are primarily determined by the accretion rate and the vertical distribution of temperature and density. In this thin region, we find that $^{56}\rm Ni$ is dominant for the flow with low accretion rate (e.g., $0.05 $$M_{\odot}$~$\rm s^{-1}$) but $^{56}\rm Fe$ is dominant for the high counterpart (e.g., $1 M_{\odot}$~$\rm s^{-1}$). The dominant $^{56}\rm Ni$ in the special region may provide a clue to understand the bumps in the optical light curve of
core-collapse supernovae.
\end{abstract}


\section{Introduction}

The popular model of the central engine in gamma-ray bursts (GRBs) is named neutrino-dominated accretion flows \cite[NDAFs, see, e.g.,][]{Popham1999,Narayan2001,Kohri2002,Di Matteo2002,Kohri2005,Lee2005, Gu2006,Chen2007,Liu2007,Liu2008,Liu2010a,Liu2010b,Liu2012,Kawanaka2007,Sun2012}. The NDAF involves a hyperaccreting spinning stellar black hole with mass accretion rates in the range of $0.01 \sim 10 M_\odot ~{\rm s}^{-1}$. In general, low (such as $0.05$ $M_{\odot}$~$\rm s^{-1}$) and high ($1$ $M_{\odot}$~$\rm s^{-1}$) mass accretion rate correspond to long ($T_{90}> 2 \rm s$) and short ($T_{90} < 2 \rm s$) duration GRBs, which are originated from collapsar \citep{Woosley1993,Paczynski1998} and the merger of two neutron stars or a neutron star and a black hole \citep{Eichler1989,Paczynski1991,Narayan1992}, respectively. The model can give reasonable explanations to the progenitor and energetics of fireball in GRBs by the neutrino annihilation or magnetohydrodynamic processes. The extreme state is a hotbed to produce heavy elements, and the central region of GRBs is an ideal location to supply an extremely hot and dense state. Actually, nucleosynthesis should also be involved in NDAFs model. \citet{Liu2007} studied the radial structure and neutrino annihilation luminosity of NDAFs. We assumed that the heaviest nucleus is $^4 \rm He$, which implies that the numerical value of the electron fraction at the radial outer boundary is 0.5. \citet{Liu2008} studied the vertical structure and luminosity of NDAFs by the above assumption. \citet{Kawanaka2007} assumed that the inflowing gas is composed primarily of neutron-rich iron group nuclei, and the electron fraction is 0.42, then studied the radial structure and stability of NDAFs for the different mass accretion rate, using a realistic equation of state \citep{Lattimer1991} in order to properly treat the dissociation of nuclei. Unfortunately, they did not provide more information on heavy nuclei. To our knowledge, the self-consistent vertical distribution of elements without making a limit for the electron fraction has not been previously investigated.

The purpose of this paper is to investigate the element distribution in the vertical direction of NDAFs with detailed neutrino physics and precise nuclear statistical equilibrium (NSE). The plan is as follows. In Section 2, with the self-similar assumption in the radial direction, we numerically solve the differential equations of NDAFs in the vertical direction with the proton-rich NSE \citep{Seitenzahl2008} and reasonable boundary condition. In Section 3, we present the vertical distribution of physical quantities, such as the density, temperature and electron fraction, and show the mass fraction of the main elements at the various radii for the different mass accretion rates. Discussion and conclusions are made in Section 4.

\section{Equations and boundary condition}

\subsection{Equations}

In order to facilitate self-similar process, we adopt the Newtonian potential $\psi = - GM/r$, where $M$ is the mass of the central black hole, and suppose the accretion flow is time-independent and axisymmetric in spherical coordinates ($r$, $\theta$, $\phi$), i.e., $\partial/\partial t =\partial/\partial \phi = 0$.

The basic equations are composed of continuity and momentum equations \citep[see e.g.,][]{Narayan1995,Xue2005}. The self-similar assumptions are adopted in the radial direction to simplify the hydrodynamic equations, and the vertical velocity is ignored ($v_\theta=0$), then we obtain the vertical hydrodynamic equations as follows \citep{Gu2009,Liu2010a,Liu2012}, \beq\ \frac{1}{2} {v_r}^2 + \frac{5}{2} {c_{\rm s}}^2 + {v_\phi}^2 - r^2 {\Omega_{\rm K}}^2 = 0, \eeq\ \beq\ \frac{1}{\rho} \frac{d p}{d \theta} = {v_\phi}^2 \cot \theta, \eeq\ \beq\ v_r = -\frac{3}{2} \frac{\alpha {c_{\rm s}}^2}{r \Omega_{\rm K}}, \eeq  where $v_r$ and $v_{\phi}$ are the radial and azimuthal components of the velocity, the sound speed $c_{\rm s}$ is defined as $c_{\rm s}^2 = p/\rho$, the Keplerian angular velocity is $\Omega_{\rm K} = (GM/r^3)^{1/2}$, and $\alpha$ is a constant viscosity parameter.

Furthermore, according the continuity equation, the mass accretion rate can be written as  \beq \dot{M} = -4 \pi r^2 \int_{\theta_0}^{\pi /2} \rho v_r \sin \theta d \theta, \eeq where $\rho$ is the density, and $\theta_0$, $\pi/2$ are the polar angle of the surface and equatorial plane, respectively.

The total pressure is expressed as the sum of four pressures: \beq\ p = p_{\rm gas} + p_{\rm rad} + p_{\rm e} + p_\nu, \eeq where $p_{\rm gas}$, $p_{\rm rad}$, $p_{\rm e}$, and $p_\nu$ are the gas pressure from nucleons, the radiation pressure of photons, the degeneracy pressure of electrons, and the radiation pressure of neutrinos, respectively. Detailed expressions of the pressure components were given in \citet{Liu2007}. Additionally, we assume the polytropic relation in the vertical direction, $p = K \rho ^{4/3}$ , where $K$ is a constant and $\gamma=4/3$ is the adiabatic index.

Considering the energetic balance, the energy equation is written as \citep{Liu2010a,Liu2012} \beq\ Q_{\rm vis} = Q_{\rm adv} + Q_\nu, \eeq where $Q_{\rm vis}$, $Q_{\rm adv}$, and $Q_\nu$ are the viscous heating, advective cooling and neutrino cooling rates per unit area, respectively. We ignore the cooling of photodisintegration of $\alpha$-particles and other heavier nuclei \citep{Liu2010a,Liu2012}. The viscous heating rate per unit volume $q_{\rm vis} = \nu \rho r^2 [\partial (v_{\phi}/r)/\partial r]^2$ and the advective cooling rate per unit volume $q_{\rm adv} = \rho v_r (\partial e/\partial r - (p/\rho^2) \partial \rho/\partial r)$ ($e$ is the internal energy per unit volume) are expressed, after self-similar simplification, as \beq\ q_{\rm vis} = \frac{9}{4} \frac{\alpha p v_{\phi}^2}{r^2 \Omega_{\rm K}}, \eeq \beq\ q_{\rm adv} = - \frac{3}{2} \frac{(p-p_{\rm e}) v_r}{r}, \eeq where the entropy of degenerate particles is neglected. Thus the vertical integration of $Q_{\rm vis}$ and $Q_{\rm adv}$ are the following \citep{Liu2010a,Liu2012}: \beq\ Q_{\rm vis} = 2 \int_{\theta_0}^{\frac{\pi}{2}} q_{\rm vis}  r \sin {\theta} d\theta \ , \eeq \beq\ Q_{\rm adv} = 2 \int_{\theta_0}^{\frac{\pi}{2}} q_{\rm adv}  r \sin {\theta} d\theta . \eeq The cooling due to the neutrino radiation $Q_\nu$ can be defined as \citep{Lee2005,Liu2012} \beq\ Q_\nu = 2 \displaystyle{\sum_{k}}  \int_{\theta_0}^{\frac{\pi}{2}} q_{\nu_k} {\rm e}^{-\tau_{\nu_k}} r \sin {\theta} d\theta \ , \eeq where $k$ represents different types of neutrinos and antineutrinos, and $q_{\nu_k}$ is the sum of the cooling rates per unit volume due to the Urca processes, electron-positron pair annihilation, nucleon-nucleon bremsstrahlung, and Plasmon decay, hereafter we represent then with $q_i$ ($i$=1, 2, 3, 4), respectively. The optical depth of neutrino $\tau_{\nu_k}$ which includes absorption $\tau_{a, {\nu_k}}$ and scattering $\tau_{s, {\nu_k}}$, is written as \beq \tau_{\nu_k}= \tau_{a, {\nu_k}} + \tau_{s, {\nu_k}}, \eeq where these two optical depth can be defined as \beq \tau_{a, {\nu_k}} \approx \displaystyle{\sum_{i}} \frac{\int_{\theta_0}^{\theta} 2 q_i r d \theta}{7 \sigma T^4}, \eeq  \beq \tau_{s, {\nu_k}} \approx \displaystyle{\sum_{j}} \int_{\theta_0}^{\theta} \sigma_j n_j  r d \theta, \eeq where $\sigma_j$ and $n_j$ ($j=1$, 2, 3, 4) are the cross sections and the number density of protons, neutrons, $\alpha$-particles, and electrons, respectively \cite[e.g.,][]{Kohri2005,Chen2007}. It is noteworthy that different type of neutrinos involve different processes \cite[for details, see][]{Liu2007}, even so, we still show the sum of all the cooling rates for any type of neutrinos or antineutrinos as Equation (13) for simplicity. We ignore the absorption and scattering reaction by the heavy nuclei in the model. In order to embody the complicated microphysics in NDAFs, we assume that the disk is optically thin to neutrinos with the mass accretion rate is less than $2 M_{\odot}~\rm s^{-1}$, which was indicated by \citet{Liu2007,Liu2012}. The electron fraction can be defined as \beq Y_{\rm e} = \frac{n_{\rm p}}{n_{\rm n}+n_{\rm p}}, \eeq where $n_{\rm p}$ and $n_{\rm n}$ are the total number density of protons and that of neutrons, which has to satisfy the the condition of electrical neutrality and the chemical potential equilibrium \cite[e.g.,][]{Liu2007}.

\cite{Liu2007} calculated the electron fraction according to the simple NSE equation, the condition of electrical neutrality and a bridging formula of electron fraction that is valid in both the optically thin and thick regimes. In this paper, we use the strict NSE equations (see Section 2.2) to replace the simple one which assumed that the heaviest nuclei is $^4 \rm He$. Meanwhile, the condition of electrical neutrality still holds, which can be written as \beq \frac {\rho Y_{\rm e}}{m_u} = n_{\rm e^{-}}-n_{\rm e^{+}},\eeq where $m_u$ is the mean mass of nucleus, and $n_{\rm e^{-}}$ and $n_{\rm e^{+}}$ are the number density of the electron and positron, which can be given by the Fermi-Dirac distribution \citep{Liu2007}.

Furthermore, if $Z_i$ and $N_i$ are defined as the number of the protons and neutrons of a nucleus $X_i$, and its number density (approximately equals the mass fraction) is $n_i$, then the electron fraction can also be written as \beq Y_{\rm e}=\sum\limits_{i} n_i Z_i/ \sum\limits_{i} n_i (Z_i+N_i). \eeq However, we assume that the disk is optically thin to neutrinos, the previous bridging formula of electron fraction can be simplified as \cite[e.g.,][]{Yuan2005,Liu2007} \beq \lg{\frac{\tilde {n}_{\rm n}}{\tilde {n}_{\rm p}}}=\frac{2 \mu_{\rm e}-Q}{k_{\rm B} T},\eeq where $\mu_{\rm e}$ and $Q$ are the chemical potential and rest-mass energy difference between neutron and proton. $\tilde {n}_{\rm n}$ and $\tilde {n}_{\rm p}$ are the number density of the free neutrons and protons, corresponding to nucleus with $Z=0$, $N=1$ and $Z=1$, $N=0$ in Equation (17).

\subsection{Nucleosynthesis}

In our calculations, the NSE equations are required to be applicable for all the variations of the electron fraction. NSE established by all nuclear reactions are in the chemical equilibrium. \cite{Seitenzahl2008} studied proton-rich material in a state of NSE, which applies to almost all the range of the electron fraction. The complicated and detailed balance has been included under the condition of the equilibrium of chemical potential. They showed that $\rm ^{56}Ni$ is favored in NSE under proton-rich conditions ($Y_{\rm e} \simeq 0.5$) being different from the case of domination by the Fe-peak nuclei with the largest binding energy per nucleon that have a proton to nucleon ratio close to the prescribed electron fraction. Particularly, the lower limit of the temperature in the NSE calculation is identified at about $2\times10^9 \rm K$ \citep{Seitenzahl2008}. If the temperature is lower than this limit, the NSE solutions will not be reliable. Therefore, in our calculations we assume that all nuclear reactions would cease when the temperature is lower than this limit. The reasons are that this limit is near the surface temperature and the components change little under this limit \citep{Seitenzahl2008}. The NSE code in proton-rich environments can be downloaded from \url{http://cococubed.asu.edu/code\_pages/nse.shtml}. In the NSE, the independent variables are the density $\rho$, temperature $T$ and electron fraction $Y_{\rm e}$, which are essential in the vertical NDAF model according to the description above.

\subsection{Boundary condition}

Naturally, if the gradient of radiation pressure is larger than the gravity, outflows may occur, so we considered that it could present a boundary condition for the accretion disk. Deviated from \citet{Liu2010a}, here a boundary condition is set to be the mechanical equilibrium, like solving the principle of Eddington luminosity. Furthermore, we can ignore the radial gradient of radial velocity and pressure at the surface of the disk. There are three forces to balance, namely gravity, radiation force and centrifugal force in non-inertial reference frame. It should be pointed out that NDAFs are extremely optical thick for photons, so the radiation pressure of photons at anywhere can be written as \beq p_{\rm rad}= \frac{1}{3} a T^4, \eeq where $a$ is the radiation constant and $T$ is the temperature of gas. The mechanical equilibrium can be written as \citep{Liu2012} \beq p_{\rm rad} \mid _{\theta=\theta_0} \sigma_{\rm T} = \frac{2GMm_u}{r^2}\rm cot \theta_0, \eeq combined with Equation (19), the surface temperature can be derived as \citep{Liu2012} \beq T \mid _{\theta=\theta_0}=(\frac{6 G M m_{u}}{a \sigma_{\rm T} r^2} \rm cot \theta_0)^{\frac{1}{4}}, \eeq where $m_{\rm u}$ is the mean mass of a nucleon, $\sigma_{\rm T}$ is the Thompson scattering cross. Since the NSE and boundary condition have been taken into account, the complete equations of our model are established.

\section{Numerical results}

Figure 1 shows the variations of the density $\rho$, temperature $T$ and electron fraction $Y_{\rm e}$ with $\theta$. The solid and dashed lines
represent the solutions at $r=10r_g$ and $100r_g$, where $r_g = 2GM/c^2$ is the Schwarzschild radius, and the thick and thin lines indicate the solutions for $\dot{M} = 0.05 M_\odot ~\rm s^{-1}$ and $1 M_\odot ~\rm s^{-1}$, respectively. The variations of the density, temperature and electron fraction are similar to the solutions in \citet{Liu2010a,Liu2012}. We marked the lower limit of the temperature $\sim 2\times10^9 \rm K$ in NSE eqautions in Figure 1(b). $\dot{M} = 0.05 M_\odot ~\rm s^{-1}$ and $\dot{M} = 1 M_\odot ~\rm s^{-1}$ correspond to $Y_{\rm e}$ around 0.49 and 0.47 near the disk surface, respectively. The half-opening angles of the disks are also similar to those in \citet{Liu2012}, which have the positive correlation with the accretion rate and radius.

Figure 2 shows the variations of the mass fraction (also approximately equals the number density) of the free neutron and proton, and main elements (include $\rm ^4 He$, $\rm ^{52}Cr$, $\rm ^{54}Cr$, $\rm ^{54}Fe$, $\rm ^{56}Fe$, $\rm ^{56}Ni$, and $\rm ^{58}Ni$, corresponding to the lines with different colors) with $\theta$ at $r=10r_g$ and $100r_g$ for $\dot{M} = 0.05 M_\odot ~\rm s^{-1}$ and $1 M_\odot ~\rm s^{-1}$ corresponding to (a)-(d). In addition, according to Equation (17), the profiles of $Y_{\rm e}$ can be indicated by Figure 2. Furthermore, the open angles in different cases correspond to the lower limit of the temperature in NSE code in Figure 1(b), which is the limit of open angle in Figure 2. We consider that the nuclear reaction will not occur under this temperature limit. $\rm ^{56}Ni$ dominates at the disk surface for $\dot{M} = 0.05 M_\odot ~\rm s^{-1}$, and $\rm ^{56}Fe$ dominates for $\dot{M} = 1 M_\odot ~\rm s^{-1}$, corresponding to $Y_{\rm e}$ around 0.49 and 0.47, respectively. Other heavy nuclei also appear in these cases. The solutions shows the proportion of the nuclear matter increases with radius for the same accretion rate. The mass fraction of $\rm ^{56}Ni$ or $\rm ^{56}Fe$ near the surface increases with radius. In the middle region, $\rm ^4 He$ is dominant for all the accretion rate. The free neutrons and protons are dominant near the equatorial plane of the disk in the hot and dense state. Most of the free protons turn into the free neutrons due to the Urca process \cite[see, e.g.,][]{Liu2007}, which causes the dominant free neutrons and the decrease of electron fraction. In the model, the accretion rate determines the density and temperature. Furthermore, the radial and vertical distribution of the density and temperature determines the electron fraction and elements distribution. In simple terms, the change of the electron fraction is inversely associated with the accretion rate and radius when the free baryons dominate.

\section{Discussion and conclusions}

\subsection{Discussion}

Supernova light curve bumps have been observed in the optical afterglow of some long-duration GRBs, which is driven by the decay of $^{56}\rm Ni$ \cite[e.g.,][]{Galama1999,Woosley2006}. How to produce massive $^{56}\rm Ni$ in supernovae accompanied GRBs is a major problem remains unsolved. The possibilities are that $^{56}\rm Ni$ is originated from the central engine transported by the outflow \cite[e.g.,][]{MacFadyen1999,MacFadyen2003,Surman2011} or the explosive burning \cite[e.g.,][]{Maeda2003,Maeda2009}. Moreover, the detection of Fe K$\alpha$ X-ray lines can play an important role in understanding the nature of GRBs \cite[e.g.,][]{Lazzati1999,Kallman2003,Gou2005,Butler2007}. The observations of some X-ray afterglows with $Beppo$-$SAX$, $ASCA$, and $Chandra$ have revealed strong Fe K$\alpha$ emission lines \cite[e.g.,][]{Piro2000}. \cite{Reeves2002} reported on an $XMM$-$Newton$ observation of the X-ray afterglow of long-duration GRB 011211. The X-ray spectrum reveals evidence for emission lines of Magnesium, Silicon, Sulphur, Argon, Calcium, and possibly Nickel, arising in enriched material with an outflow velocity of order $\rm 0.1c$. Nevertheless, there are no further observations on the metal lines in GRBs since the launch of $Swift$. There are scarcely any observational evidences on Fe K$\alpha$ lines appearing in the X-ray afterglow of short-duration GRBs. Recently, many works have focused on how the heavy nuclei are produced by the central engine of GRBs. \cite{Surman2011} studied the nucleosynthesis, particularly for the progenitor of $^{56}\rm Ni$ in the hot outflows from GRB accretion disks. \cite{Metzger2011} suggested that the composition of ultrahigh energy cosmic rays becomes dominated by heavy nuclei at high energies forming GRBs jet or outflows \cite[also see][]{Sigl1995,Horiuchi2012}. \cite{Metzger2012} investigated the steady-state models of accretion disks produced by the tidal disruption of a white dwarf by a neutron star or a stellar black hole, which can produce heavier elements via burning the initial material of the white dwarf. They discussed the recently discovered subluminous Type I supernovae result from those mergers.

When a massive star collapses to a black hole, powerful supernova occurs. The newborn hyperaccreting black hole may power a GRB. The optical light curve bumps of the supernovae accompanied with GRBs is driven by the decay of $^{56}\rm Ni$ in the outflows coming from the central engine. We have described self-consistently how to produce $^{56}\rm Ni$ and other elements in the central region of GRBs with the NDAF model. Only in the low accretion rate exactly corresponding to long GRBs, $^{56}\rm Ni$ dominates near the disk surface. More daringly, if the outflow occurs from the disk surface, which consists of $^{56}\rm Ni$ and other heavy nuclei, the bumps in supernova light curve can be naturally generated due to $^{56}\rm Ni$ decay in the outflow from NDAFs. Actually, \cite{Liu2012} revisited the vertical structure of NDAFs and showed that the possible outflow may appear in the outer region of the disk according to the calculations of the vertical distribution of the Bernoulli parameter. Furthermore, We noticed that the gradient of radiation pressure may be larger than the gravity in the region out of the boundary, which has been represented in Section 2.3. Part of the material near the surface would become outflow pushed out by the radiation pressure. We also noticed that the description of NDAF is similar to that of slim disk. The high-speed outflow may be generated near the disk surface in the two-dimensional simulations of supercritical disk \citep[e.g.,][]{Ohsuga2005,Ohsuga2007,Ohsuga2011}. All these suggest that the outflow near the disk surface may appear in NDAF model.

Nickel existed in the outflows from the surface of NDAF with the low mass accretion rate may be the most important source of the long-duration GRB $\rm ^{56} Ni$ production. Analogously, iron and other heavy nuclei may also form the outflow injecting into external environment in GRBs. The NDAF model with outflows is necessary to be constructed for further theoretical explanation of the bumps in the optical light curve of core-collapse supernovae.

\subsection{Conclusions}

In this paper, we revisit the vertical elements distribution of NDAFs around black hole in spherical coordinates with detailed neutrino physics and precise nuclear statistical equilibrium and discuss the vertical elements distribution in the solutions. The major points are as follows:

\begin{enumerate}
\item
The heavy elements distribute near the surface of the disk and the free protons and neutrons dominate near the equatorial plane. The distribution of the density and temperature determine the nuclear matter distribution.
\item
In the region near the surface of NDAFs, Nickel will be dominant for low mass accretion rates, whereas iron will be dominant for high accretion rates. The dominant $^{56}\rm Ni$ may provide a clue to understand the bumps in the optical light curve of core-collapse supernovae.
\end{enumerate}

\acknowledgments

We thank the anonymous referee for a careful and constructive review, and Xue-Feng Wu and Shu-Jin Hou for beneficial discussion. TL thanks Ang Li for being invited to visit RIKEN. This work was supported by the National Basic Research Program (973 Program) of China under grant 2009CB824800, the National Natural Science Foundation of China under grants 10833002, 11003016, 11073015, 11103015, 11222328, and 11233006, and the Natural Science Foundation of Fujian Province of China under grant 2010J01017.

\clearpage

\begin{figure*}
\centering
\includegraphics[angle=0,scale=0.35]{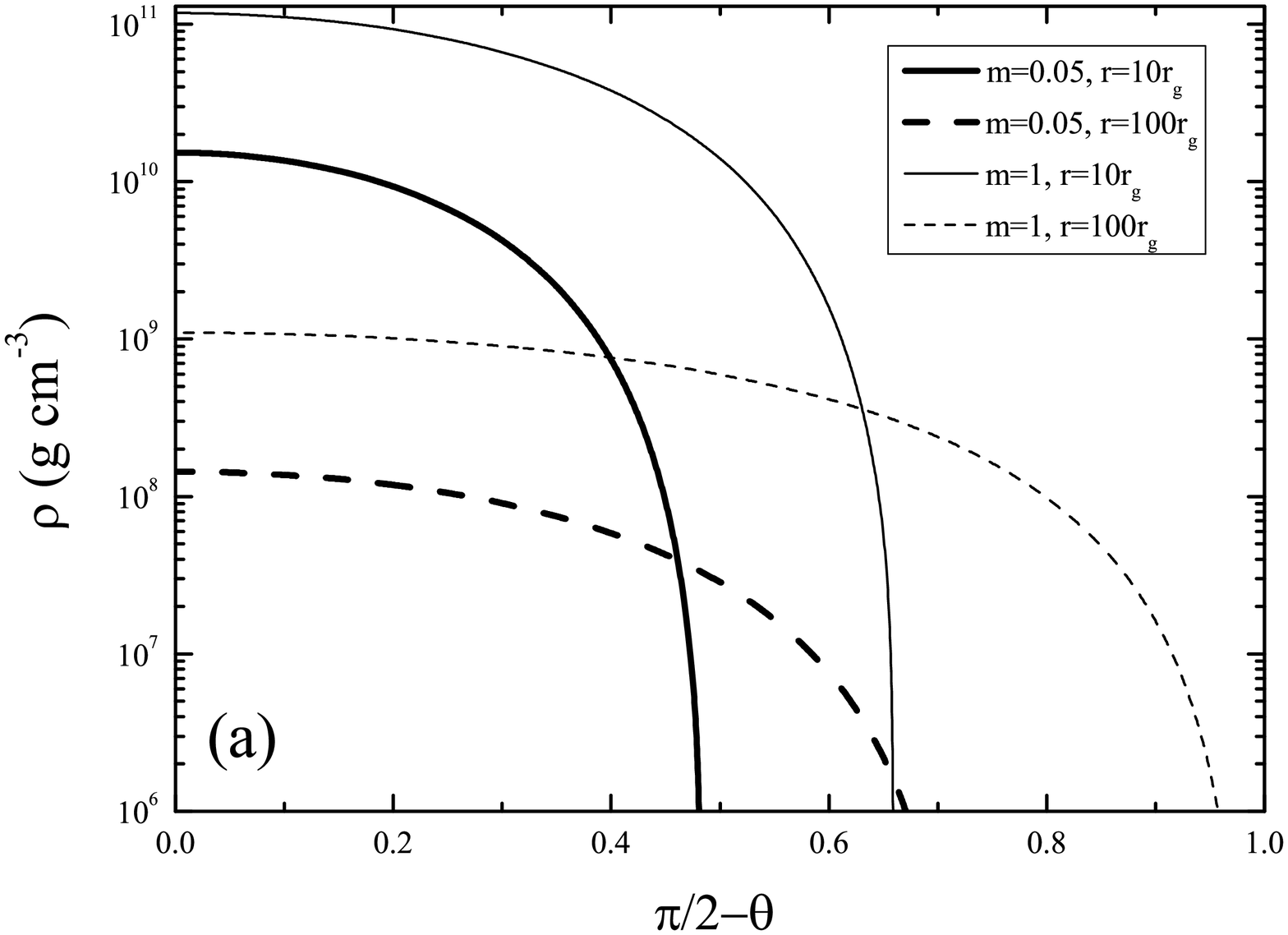}
\includegraphics[angle=0,scale=0.35]{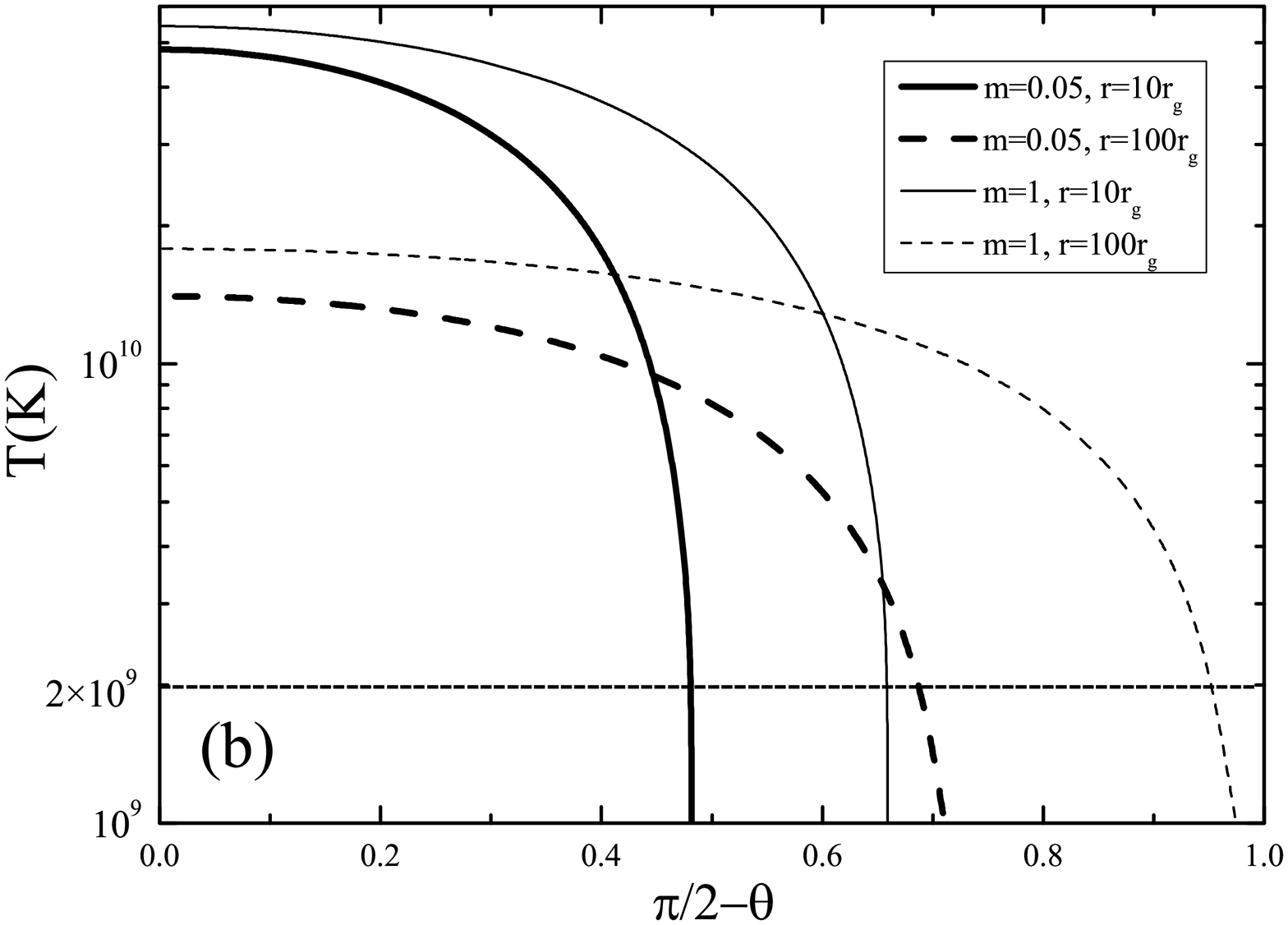}
\includegraphics[angle=0,scale=0.35]{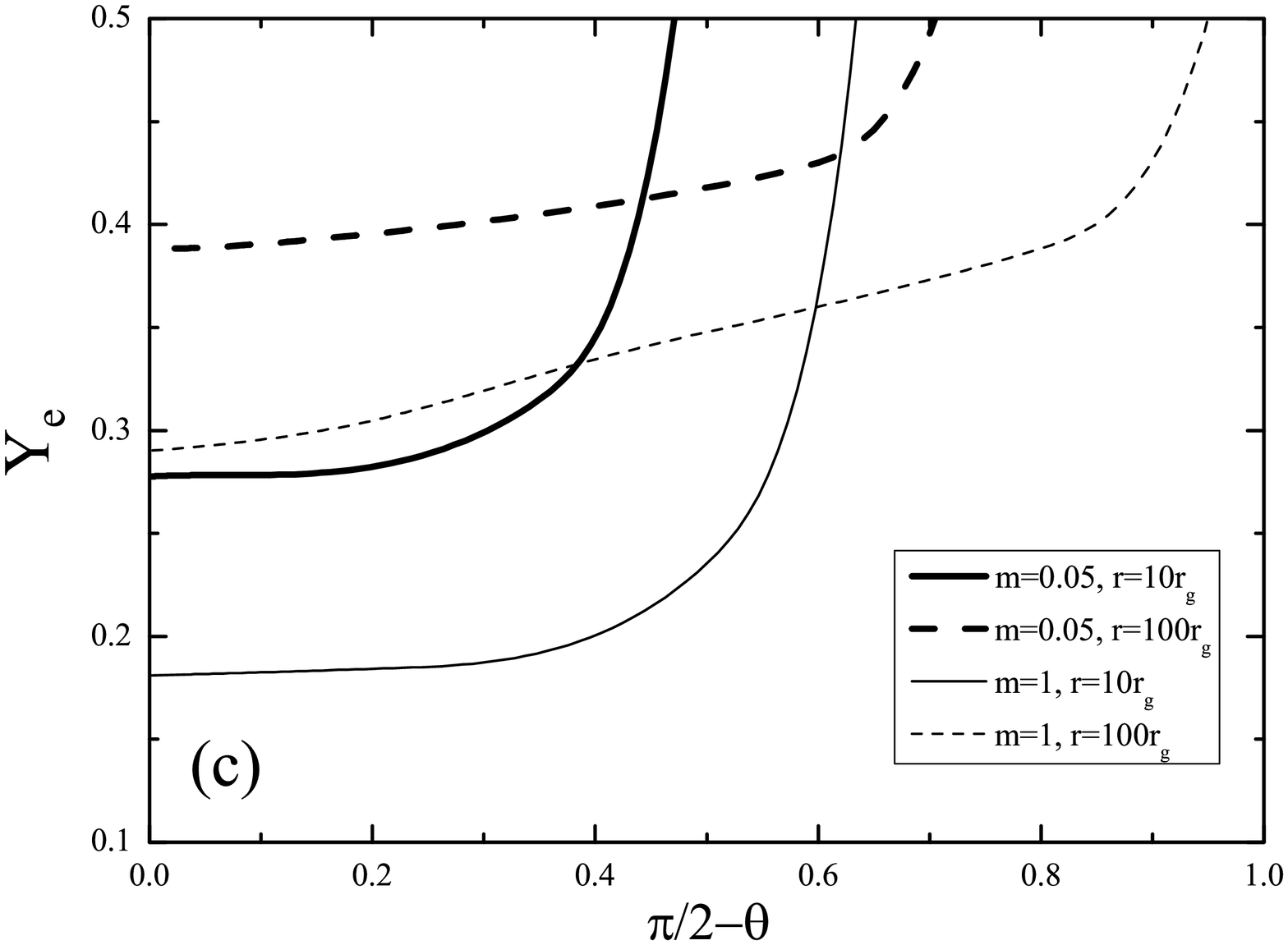}
\caption{Variations of the density $\rho$, temperature $T$ and electron fraction $Y_{\rm e}$ with
$\theta$ at $r=10r_g$ (solid lines) and $100r_g$ (dashed lines)
for $\dot{M} = m M_\odot~\rm s^{-1}$, where $m=0.05$ (thick lines) and
$1$ (thin lines).}
\label{sample-figure1}
\end{figure*}

\clearpage

\begin{figure*}
\centering
\includegraphics[angle=0,scale=0.27]{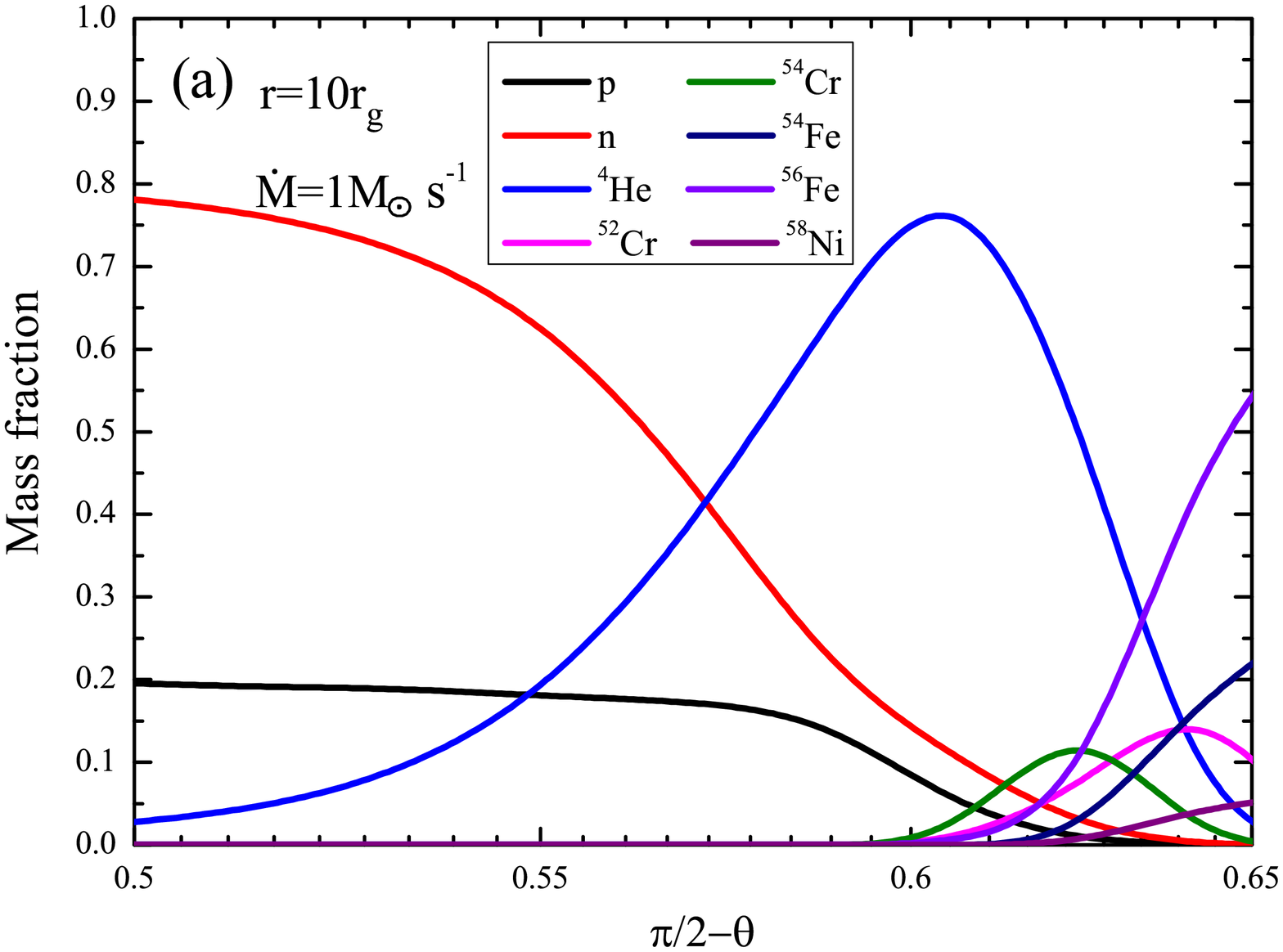}
\includegraphics[angle=0,scale=0.27]{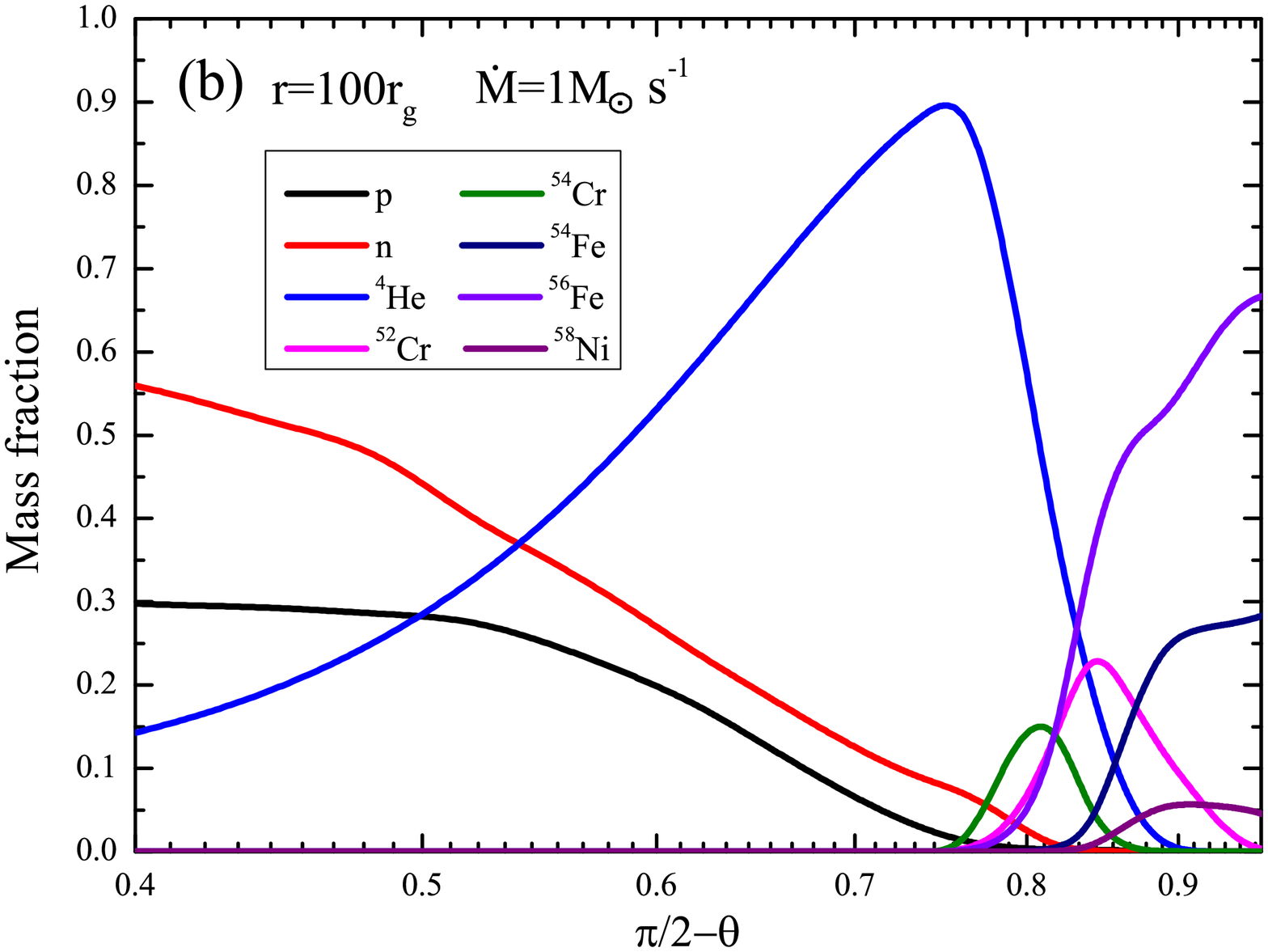}
\includegraphics[angle=0,scale=0.27]{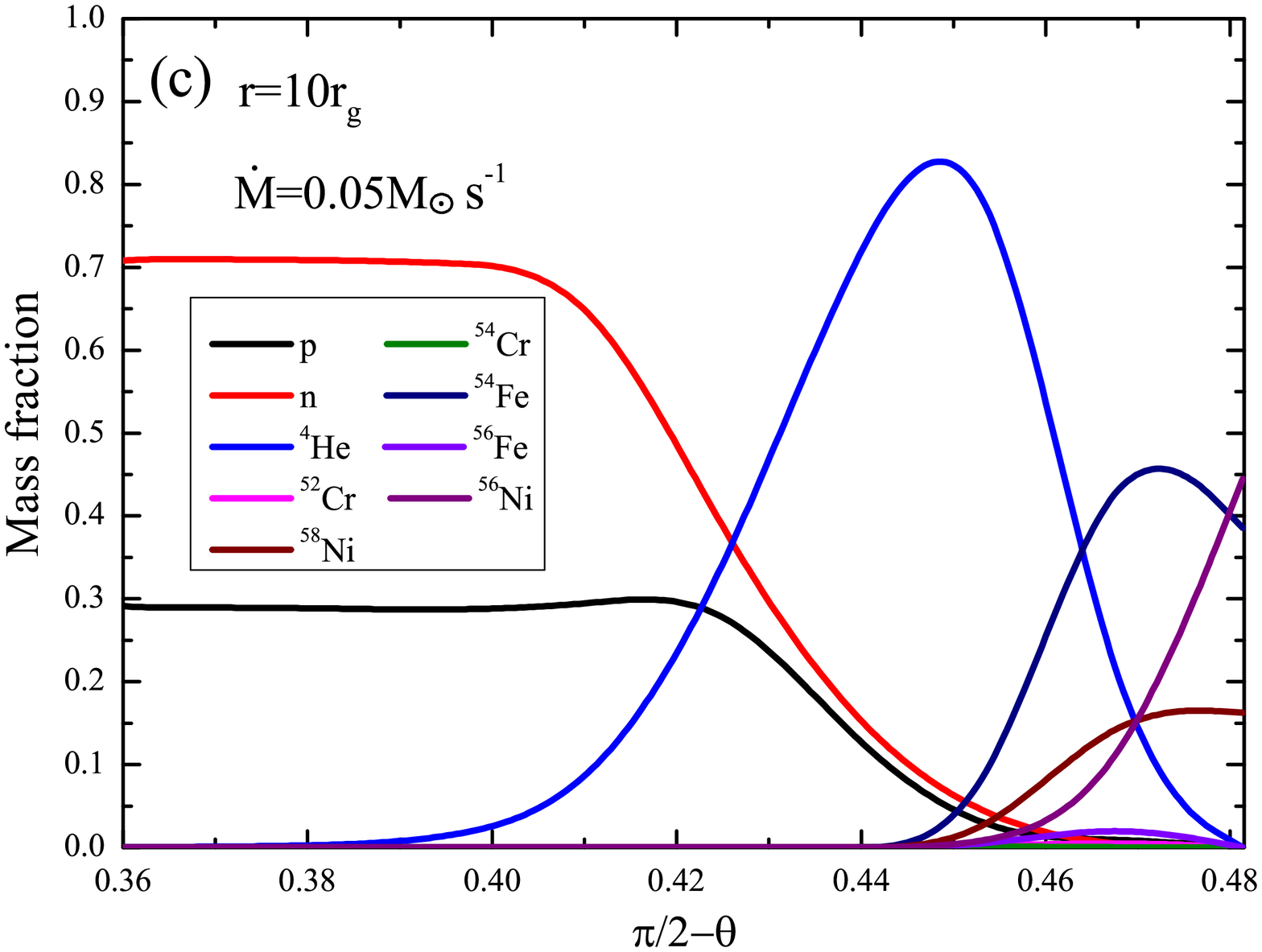}
\includegraphics[angle=0,scale=0.27]{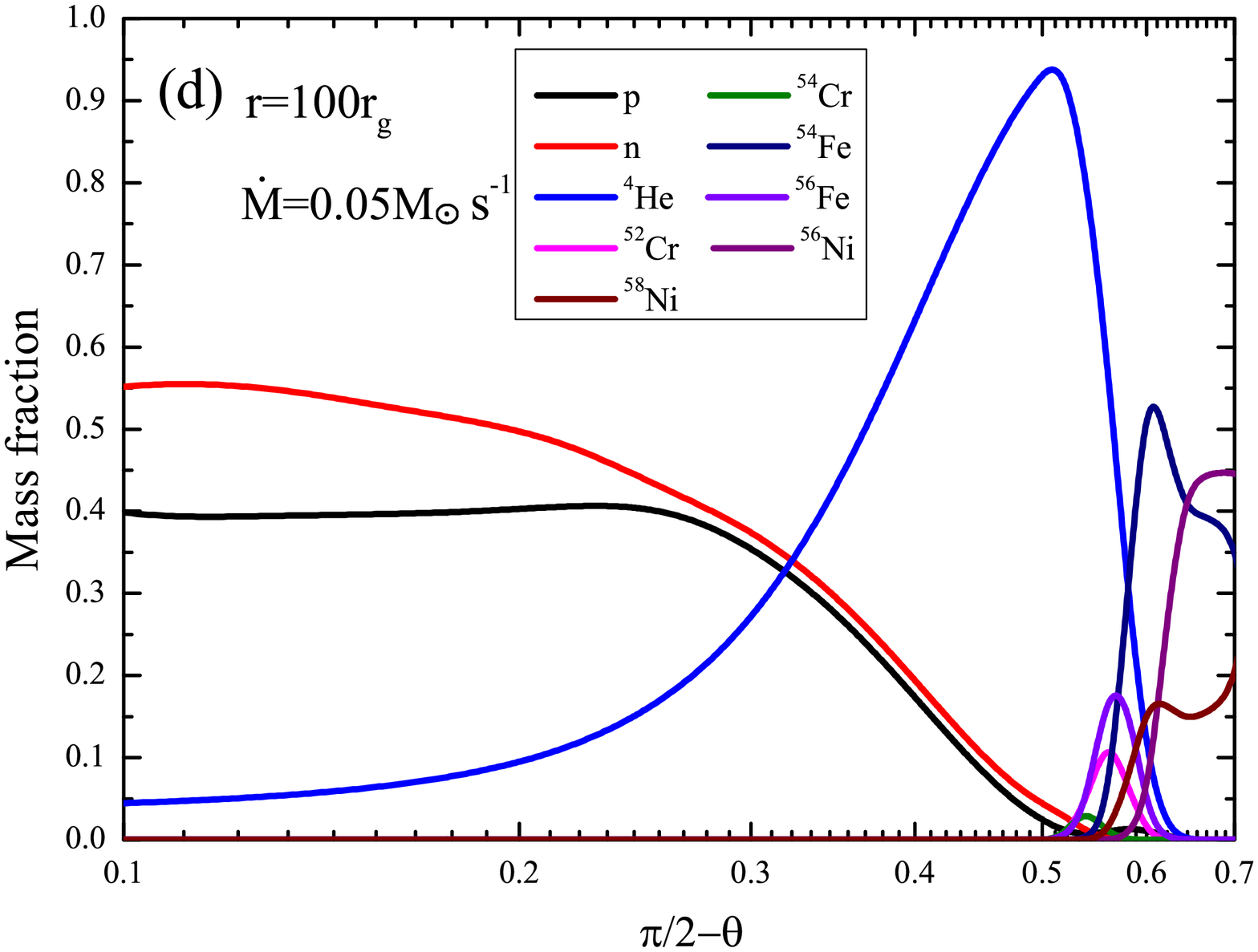}
\caption{Variations of the mass fraction of the main elements with
$\theta$ at $r=10r_g$ and $100r_g$ for $\dot{M} = 0.05 M_\odot ~\rm s^{-1}$ and $1 M_\odot ~\rm s^{-1}$.}
\label{sample-figure2}
\end{figure*}

\clearpage

\end{document}